\documentclass[sigconf,nonacm,pbalance]{acmart}
\setcopyright{none}
\usepackage{amsmath}
\usepackage{array,booktabs,tabularx}
\usepackage{xurl}
\hypersetup{colorlinks=true,linkcolor=blue,citecolor=blue,urlcolor=blue}
\newcolumntype{Y}{>{\raggedright\arraybackslash}X}

\usepackage{tikz}
\usetikzlibrary{arrows.meta}

\newcommand{\DiagnosticEpisodes}{180}
\newcommand{\CalibrationEpisodes}{294}

\newcommand{\StructuredSuccess}{24}

\newcommand{\StructuredLongSuccess}{34}
\newcommand{\EfficientShortSuccess}{35}
\newcommand{\EfficientLongSuccess}{36}
\newcommand{\AckObsolete}{89}
\newcommand{\AckGaps}{24}
\newcommand{\UnauthorizedAdoptions}{18}
\newcommand{\ConstraintAttempts}{1533}
\newcommand{\NoRelayUninformed}{375}
\newcommand{\LocalModelName}{qwen3:4b-instruct}

\newcommand{\InterventionEpisodes}{16}
\newcommand{\InterventionSuccess}{4}
\newcommand{\InterventionBaselineSuccess}{1}

\newcommand{\CalibrationProbeTotal}{9}

\begin{document}
\title[When the Plan Changes]{When the Plan Changes: \texorpdfstring{\\}{ }A Strategic World Testbed for Dynamic Multi-Agent Coordination}
\author{Joon-Seok Kim}
\orcid{0000-0001-9963-6698}
\affiliation{%
  \institution{Emory University}
  \city{Atlanta}
  \state{Georgia}
  \country{USA}
}
\email{joonseok.kim@emory.edu}

\author{Eugene Agichtein}
\orcid{0000-0002-3148-5448}
\affiliation{%
  \institution{Emory University}
  \city{Atlanta}
  \state{Georgia}
  \country{USA}
}
\email{yagicht@emory.edu}

\author{Kai Shu}
\orcid{0000-0002-6043-1764}
\affiliation{%
  \institution{Emory University}
  \city{Atlanta}
  \state{Georgia}
  \country{USA}
}
\email{kai.shu@emory.edu}

\author{Liang Zhao}
\orcid{0000-0002-2648-9989}
\affiliation{%
  \institution{Emory University}
  \city{Atlanta}
  \state{Georgia}
  \country{USA}
}
\email{liang.zhao@emory.edu}

\author{Andreas Z{\"u}fle}
\orcid{0000-0001-7001-4123}
\affiliation{%
  \institution{Emory University}
  \city{Atlanta}
  \state{Georgia}
  \country{USA}
}
\email{azufle@emory.edu}

\renewcommand{\shortauthors}{Kim et al.}

\begin{abstract}
Changing an objective does not automatically revise agents' earlier plans or delegated tasks. Acknowledgment can coexist with obsolete execution, while misleading instructions can redirect work or discard retained constraints. We present a Strategic World simulation testbed that separates authoritative objectives, local plans, notice receipt, and consequential action. Six intervention families vary objectives, conditions, authority, membership, and communication. Matched protocols and deadlines establish feasibility for the tested disturbances. Controlled trajectories demonstrate that identical completion and acknowledgment summaries can conceal uninformed obsolete work, obsolete work after receipt, or blocked unsafe intent. Local language-model trials and instructed action probes reveal competence limits that constrain interpretation of dynamic failures. A further \InterventionEpisodes{}-episode comparison evaluates an engineered guard that prioritizes pending notice submissions over local work. It completes \InterventionSuccess{}/4 held-out settings, compared with \InterventionBaselineSuccess{}/4 for a calibrated assisted baseline. Each episode contains one large language model actor and two fixed peers. These results establish diagnostic distinctions and bounded policy feasibility under explicit simulation assumptions; they do not establish frontier-model reliability or autonomous team adaptation.

\end{abstract}
\maketitle
\section{Introduction}
An artificial intelligence (AI) agent executes within a history of decisions. It may have stored a plan, delegated work, or initiated a process whose effects occur later. Classical accounts of teamwork make commitments and communication about shared activity central to coordinated action~\cite{cohen1991teamwork,grosz1996sharedplans}. Recent studies of intent drift and dynamic goal revision show that large language model (LLM) agents can retain superseded requirements or satisfy only part of a revised request~\cite{zhang2026intentflux,zhang2026dynarev}. Adapting a team therefore requires updating both the intended objective and the commitments already distributed among its members. At the same time, agents must distinguish legitimate revisions from untrusted instructions: tool-mediated prompt injection can redirect execution toward an attacker's task~\cite{debenedetti2024agentdojo}.

Consider two agents assigned to repair a fictional water network. An authorized update changes their priority to delivering bottled water, while preserving a restriction against using a protected route. Agent \textit{A} receives the update and replies, ``Understood,'' but spends its next turn continuing the repair instead of notifying Agent \textit{B}. Agent \textit{B}, disconnected from communication, also continues repairing. The visible actions are alike, yet their causes differ: \textit{A} has an opportunity to change its behavior, whereas \textit{B} lacks the update. If both later finish the delivery, final success conceals the obsolete work. A misleading message claiming that the protected route is now permitted creates a further test: adopting the new delivery objective must not erase a retained constraint.

The difficulty is organizational as well as individual. Shared assignments persist when teammates miss messages, leave, or join with an earlier task description. Communication and individual task competence can make different contributions to team performance, as demonstrated in long-horizon coordination experiments~\cite{tessera2026alem}. Analytical benchmarks already measure incremental progress beyond final success, and trace-based taxonomies distinguish multiple sources of multi-agent failure~\cite{ma2024agentboard,cemri2025failures}. The remaining measurement problem addressed here is to relate each consequential action to the authorized objective, retained obligations, and information available to its actor. Receipt, acknowledgment, plan adoption, and subsequent work must be distinguished to separate missing information from behavior after an update.

We develop a testbed, an executable simulation environment for controlled experiments, using Strategic World, a graph-structured simulation model that links strategic objectives, operational missions, and tactical actions (Section~\ref{sec:strategic_world})\footnote{GitHub repository will be made public upon manuscript acceptance.}. The testbed separates authoritative world state, local plans, and observations so that a revision can affect the organization without immediately changing every participant's behavior. We instantiate a bounded operational workflow to study these coordination mechanisms under controlled interventions.

The contribution comprises (1) a state and intervention model that separates authoritative change from distributed adaptation, (2) an executable workflow implementation and six controlled scenario families, and (3) an evaluation separating informed obsolete work, uninformed work, unsafe attempts, and verified outcomes. We evaluate scripted controls with finite internal state designed to expose specific failure mechanisms and run local language-model pilots through the same interfaces. Matched local protocols and deadlines establish feasibility for the tested disturbances. Constructed execution histories, called trajectories, show that completion and acknowledgment can coincide despite different failure mechanisms. Language-model episodes with fixed objectives and membership (static episodes) and isolated action probes separate basic interface competence from interpretation of dynamic failures. A further controlled experiment uses a calibrated assisted baseline to test an execution guard motivated by trace diagnosis under timing and workload conditions reserved for evaluation (held-out conditions). The results concern measurement and bounded policy behavior, rather than a comparative ranking of the most capable contemporary models.

Our evaluation asks four questions: (1) how protocol and deadline choices affect completion under matched disturbances; (2) which distinctions between missing information, post-receipt behavior, and blocked unsafe intent disappear under completion and acknowledgment summaries; (3) whether independent language-model actors complete static workflows and how their behavior changes under revised objectives or misleading direction; and (4) whether a trace-informed execution intervention improves held-out completion relative to a calibrated baseline and nonspecific assistance. Observation isolation, which keeps inaccessible state out of policy inputs, and deterministic replay, which reconstructs a trajectory from recorded actions, are prerequisites for these comparisons. The experiments characterize designed conditions and a small local-model sample; population-level claims about frontier models require separate sampling and statistical analysis.

\section{Related Work}
\paragraph{Revising objectives while preserving commitments.}
Collaborative action depends on commitments that persist beyond an individual decision. Joint-intention and shared-plan accounts make communication about a common activity, partial knowledge, and delegated action central to teamwork~\cite{cohen1991teamwork,grosz1996sharedplans,tambe1997teamwork}. Controlled evaluations of evolving instructions document interference from superseded requirements and incomplete compliance with revised requests~\cite{zhang2026intentflux,zhang2026dynarev}. Together, these perspectives suggest that revising an objective involves both interpreting the new request and reconciling commitments created under the old one.

For a team, that reconciliation is distributed. Revising one actor's plan need not notify its peers or replace their assignments, and some obligations must survive the change. The present study connects goal revision to these persistent organizational effects by separating authoritative objectives, actor-local plans, and ongoing work. Its objective labels and independent assignments provide a bounded experimental abstraction of commitments rather than a full implementation of shared-plan semantics.

\paragraph{Steerability, authority, and memory.}
Behavioral redirection depends on both the mechanism of control and whether the direction should be followed. Instruction privilege and prompt-injection research address how trusted instructions can be distinguished from untrusted content~\cite{wallace2024hierarchy,debenedetti2024agentdojo}. Activation steering intervenes more directly in the intermediate numerical states computed during model execution; such interventions can support targeted behavioral control or induce adversarial behavior~\cite{rimsky2024steering,wang2024trojan}. Contextual-privacy results further show that representing a norm internally does not guarantee compliance with that norm in generated behavior~\cite{wang2026privacy}. Behavioral acknowledgment, internal representation, and consequential compliance therefore require different evidence.

Memory introduces another distinction between retaining information and making it guide action. Agent-memory research separates storage, cognitive role, and the subject of remembered experience, while adaptive indexing changes future access without rewriting stored content~\cite{huang2026memory,hu2026amie}. These mechanisms are relevant to changing instructions because an old objective can remain stored even when it should cease to govern execution. Our study measures observable responses to notices and controller rules, with explicit separation of local objectives, conversational history, and assignments. It does not probe model representations or identify a memory-retrieval mechanism. Authority is likewise represented as an experimental condition: origin verification and revision permission are distinct concepts, although the evaluated legitimate and forged notices vary them together.

\paragraph{Simulating organized interaction.}
The organization of a multi-agent system depends jointly on who communicates, what memory is shared, and how state changes over time~\cite{song2026networks}. Language-mediated simulations demonstrate how memory, planning, and grounded interaction can support individual and emergent social behavior~\cite{park2023generative,vezhnevets2023concordia}. Coordination benchmarks impose complementary demands through milestones, communication topologies, specialization, and long-horizon joint tasks~\cite{zhu2025multiagentbench,tessera2026alem}. These approaches make different aspects of interaction observable; organizational topology is distinct from a simulation's strategic, operational, and tactical decision levels.

Changing participants adds a further challenge. Ad hoc teamwork concerns cooperation without prior coordination~\cite{stone2010adhoc}, whereas the present experiments begin with prescribed actors and assignments and vary their subsequent availability. Strategic World supplies a controlled setting in which assignments persist through delayed communication and departure. Explicit state transitions make ownership, delivery, progress, and blocked attempts inspectable, enabling comparisons under matched disturbances. This abstraction supports analysis of adaptation within an existing organization, without establishing social realism or the ability to form a new team autonomously.

\paragraph{Measuring adaptation under partial knowledge.}
Process evaluation already extends beyond final success: incremental progress and milestone measures expose intermediate achievement, while failure taxonomies distinguish design, alignment, and verification problems~\cite{ma2024agentboard,zhu2025multiagentbench,cemri2025failures}. For changing objectives, the additional issue is which information an actor could use when it acted. Event-ordering and distributed-snapshot research formalize why local histories and global state must be distinguished~\cite{lamport1978time,chandy1985snapshots}. Issuing a revision, receiving it, and acting after receipt are consequently different events, even when final completion is identical.

Our diagnostic contribution combines these process and information perspectives. Conditioning work on notice receipt separates uninformed obsolete execution from behavior after an update; recording blocked attempts distinguishes policy intent from an enforced safeguard. Reporting eligible actors and responses not observed before departure or the deadline avoids equating a latency among observed responders with team-wide adaptation. Constructed trajectory comparisons establish distinctions lost by completion and acknowledgment summaries, and a subsequent controlled experiment evaluates a trace-motivated intervention. Durable message acceptance and serialized ownership are supplied by the simulator, so these comparisons concern behavioral coordination under explicit guarantees rather than distributed agreement achieved by the policies.

\section{Strategic World Model}
\label{sec:strategic_world}
Agent-based modeling examines system behavior through the actions and interactions of autonomous entities~\cite{macal2010tutorial}. Strategic World represents a multi-agent environment as a graph of capabilities and their dependencies. A capability describes a class of state and behavior, such as mission assignment, communication, infrastructure service, or unit movement. Actors are consequential entities in the simulated world. They instantiate these capabilities and select actions through policies, rules or model-based procedures that map local observations to action choices. A versioned profile selects capability implementations, a versioned mode defines roles and evaluation rules, and a scenario supplies concrete actors, tasks, and disturbances. A graph node therefore denotes a modeling responsibility, not necessarily an individual agent or language-model process.

Wargaming guidance covers national strategic, strategic, operational, and tactical questions~\cite{mod2017wargaming}. Strategic World uses a related organization with four domains of decision-making, including an explicit geopolitical domain for relationships among actors. The \emph{geopolitical} level represents actors, relationships, coalitions, and agreements. The \emph{national strategic} level represents policy objectives, institutions, resources, and escalation constraints. The \emph{operational} level connects these objectives to missions, logistics, communication, intelligence, infrastructure, and civilian services. The \emph{tactical} level represents units, local observations, movement, and compliance with assignments. \emph{Cross-cutting services} supply scheduling, policy execution, events, evaluation, replay, and safeguards across these domains.

Linking representations at different resolutions introduces consistency and calibration problems, as studied in multiresolution modeling~\cite{davis1998experiments}. Here, decision domains organize responsibilities, while graph edges specify dependencies and permissible effects. They do not by themselves define aggregation or disaggregation between resolutions. The experiments in this paper evaluate a bounded operational workflow within that graph.

\begin{figure}[t]
\centering
\includegraphics[width=\linewidth]{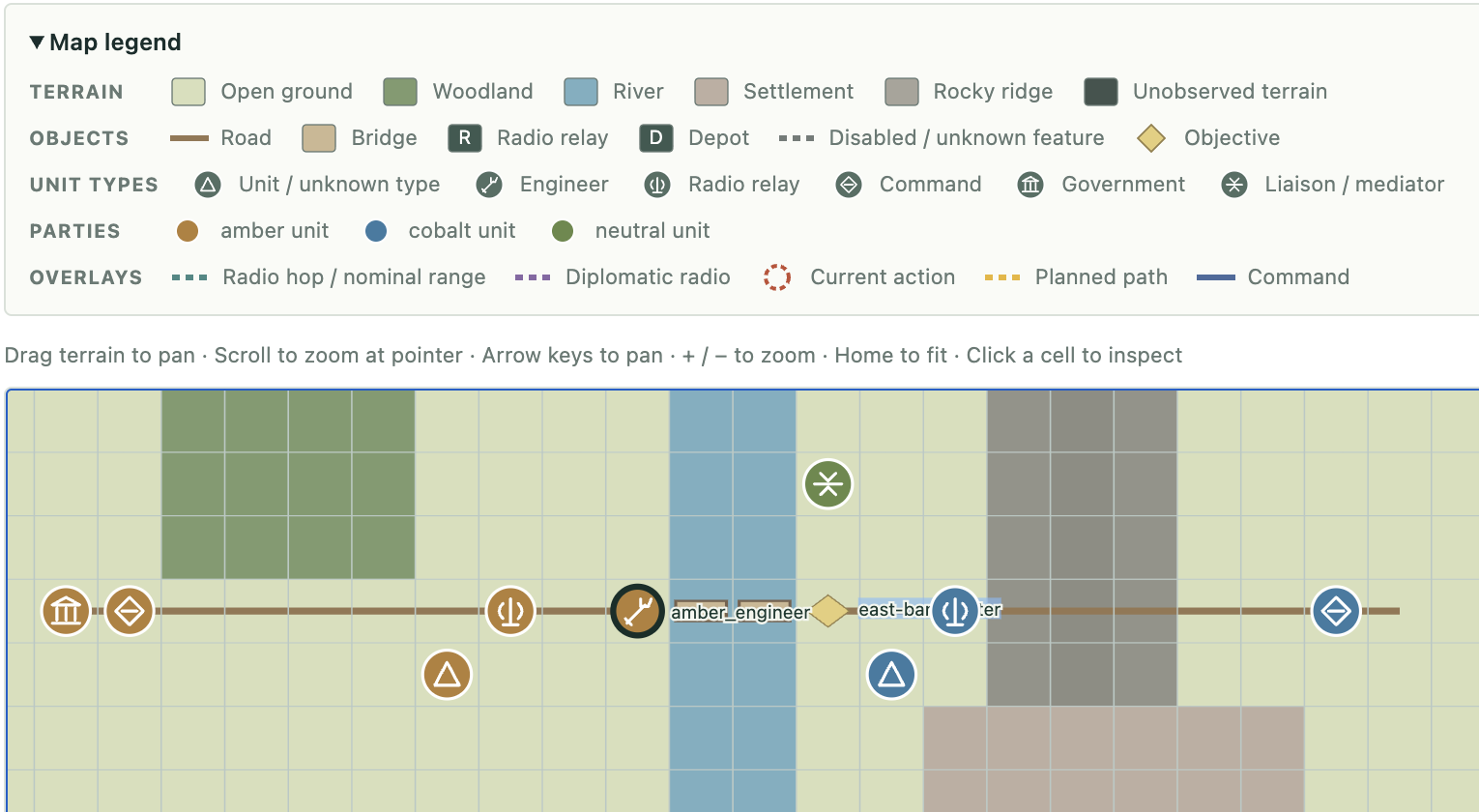}
\caption{Geographic atlas of a fictional relief scenario in Strategic World, showing terrain, infrastructure, unit types, and communication overlays.}
\Description{An atlas interface with a terrain and object legend, a river crossed by a road, infrastructure markers, and unit icons.}
\label{fig:atlas}
\end{figure}

\begin{figure*}[t]
\centering
\includegraphics[width=0.7\textwidth]{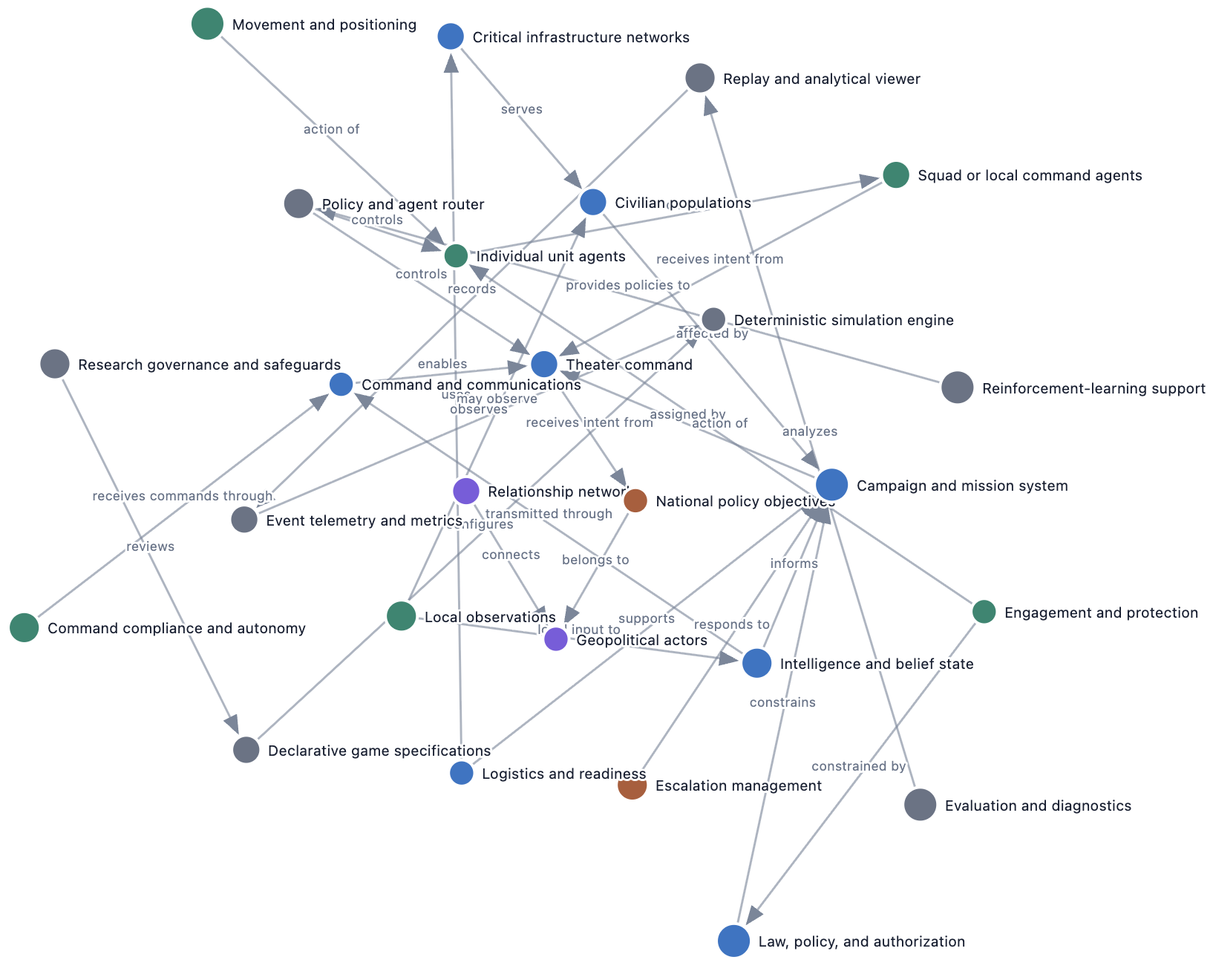}
\caption{Selected capabilities and relationships in Strategic World. Purple nodes denote geopolitical capabilities, brown national strategic capabilities, blue operational capabilities, green tactical capabilities, and gray cross-cutting services. Arrows denote the labeled semantic relationships, including dependencies, authority, observation, and evaluation; they are not a chronological message-flow diagram. The coordination experiments instantiate the operational assignment--communication--execution path within this broader model.}
\Description{A simplified directed capability graph with five color-coded categories. National policy connects to theater command and missions; communication links command and unit compliance; local observations connect units to intelligence. Infrastructure, civilian populations, logistics, authorization, and escalation constrain or receive effects from operational decisions. Simulation, policy routing, telemetry, replay, evaluation, and safeguards form cross-cutting services.}
\label{fig:model_graph}
\end{figure*}

Figure~\ref{fig:model_graph} shows selected capabilities and their labeled relationships. For example, theater command receives national intent, communication enables command, and unit compliance depends on delivered messages. Unit actions can affect infrastructure and civilian services, while observations inform operational beliefs. These relations specify dependencies and permissible effects; they do not imply that information is instantly shared or that every team has a single coordinator.

The simulation maintains authoritative world state while each actor observes only an accessible projection and retains its own plan. This distinction supports persistent consequences and asynchronous knowledge: an objective may change nationally while an operational assignment or a unit's local task remains unchanged. Coordination structure is independent of the model's decision levels. Peer agents can execute a mission without a designated leader while remaining subject to common objectives, resource constraints, and authorization rules.

\section{Testbed Design}
\label{sec:testbed}
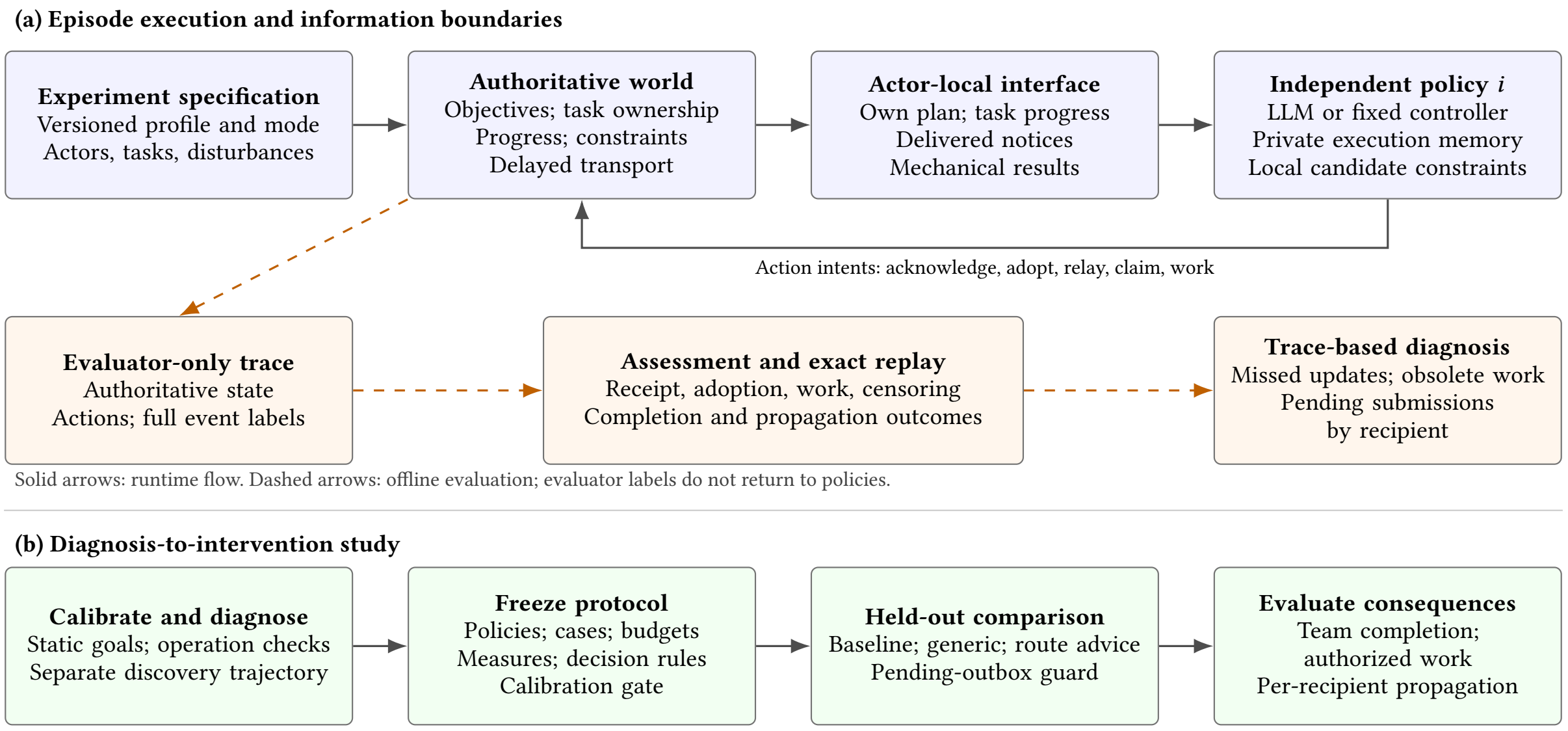
\begin{figure*}[t]
\centering
\begin{tikzpicture}[
  x=1cm,y=1cm,
  box/.style={draw=black!55,rounded corners=2pt,align=center,text width=3.25cm,minimum height=1.22cm,inner sep=4pt,font=\footnotesize},
  runtime/.style={box,fill=blue!5,minimum height=1.5cm},
  audit/.style={box,fill=orange!7,minimum height=1.5cm},
  study/.style={box,fill=green!5,minimum height=1.6cm},
  flow/.style={-{Latex[length=2mm]},semithick,draw=black!70},
  record/.style={-{Latex[length=2mm]},semithick,dashed,draw=orange!75!black}
]
\node[anchor=west,font=\footnotesize\bfseries] at (-1.78,1.06) {(a) Episode execution and information boundaries};
\node[runtime] (spec) at (0,0) {\textbf{Experiment specification}\\Versioned profile and mode\\Actors, tasks, disturbances};
\node[runtime] (world) at (4.1,0) {\textbf{Authoritative world}\\Objectives; task ownership\\Progress; constraints\\Delayed transport};
\node[runtime] (local) at (8.2,0) {\textbf{Actor-local interface}\\Own plan; task progress\\Delivered notices\\Mechanical results};
\node[runtime] (policy) at (12.3,0) {\textbf{Independent policy $i$}\\LLM or fixed controller\\Private execution memory\\Local candidate constraints};
\draw[flow] (spec) -- (world);
\draw[flow] (world) -- (local);
\draw[flow] (local) -- (policy);
\draw[flow] (policy.south) -- (12.3,-1.25) -- node[below,font=\scriptsize,align=center] {Action intents: acknowledge, adopt, relay, claim, work} (4.1,-1.25) -- (world.south);
\node[audit,text width=3.25cm] (trace) at (0,-2.7) {\textbf{Evaluator-only trace}\\Authoritative state\\Actions; full event labels};
\node[audit,text width=4.6cm] (assessment) at (6.15,-2.7) {\textbf{Assessment and exact replay}\\Receipt, adoption, work, censoring\\Completion and propagation outcomes};
\node[audit,text width=3.25cm] (diagnosis) at (12.3,-2.7) {\textbf{Trace-based diagnosis}\\Missed updates; obsolete work\\Pending submissions\\by recipient};
\draw[record] (world.south west) -- (trace.north);
\draw[record] (trace) -- (assessment);
\draw[record] (assessment) -- (diagnosis);
\node[font=\scriptsize,anchor=west,text=black!75] at (-1.78,-3.62) {Solid arrows: runtime flow. Dashed arrows: offline evaluation; evaluator labels do not return to policies.};
\draw[black!20] (-1.78,-3.92) -- (14.08,-3.92);
\node[anchor=west,font=\footnotesize\bfseries] at (-1.78,-4.28) {(b) Diagnosis-to-intervention study};
\node[study] (discovery) at (0,-5.3) {\textbf{Calibrate and diagnose}\\Static goals; operation checks\\Separate discovery trajectory};
\node[study] (freeze) at (4.1,-5.3) {\textbf{Freeze protocol}\\Policies; cases; budgets\\Measures; decision rules\\Calibration gate};
\node[study] (heldout) at (8.2,-5.3) {\textbf{Held-out comparison}\\Baseline; generic; route advice\\Pending-outbox guard};
\node[study] (result) at (12.3,-5.3) {\textbf{Evaluate consequences}\\Team completion; authorized work\\Per-recipient propagation};
\draw[flow] (discovery) -- (freeze);
\draw[flow] (freeze) -- (heldout);
\draw[flow] (heldout) -- (result);
\end{tikzpicture}
\caption{Overall testbed design. Each actor receives a local projection and selects action intents; the simulator commits their consequences and delivers due notices. Full state and compatibility classifications are retained for offline assessment, not policy feedback. The study uses calibration and a separate discovery trajectory to select an intervention, freezes the protocol, and evaluates matched held-out settings. Actor-local assistance is an explicit policy component, not a change to world legality.}
\Description{Two-panel architecture diagram. The upper panel connects a versioned experiment specification to authoritative world state, actor-local observations and mechanical feedback, and independent policies. Policy actions return to the world. A separate dashed path leads from world records to offline assessment, replay, and diagnosis. The lower panel shows calibration and discovery, protocol freeze, held-out policy comparison, and evaluation of completion and propagation. No evaluator-only classification feeds back to an executing actor.}
\label{fig:overall_design}
\end{figure*}

Figure~\ref{fig:overall_design} summarizes the execution loop, its information boundaries, and the experimental workflow. Runtime policies receive actor-local observations and mechanical outcomes: whether their own previous action progressed, conflicted, or was blocked, without an evaluator label indicating compatibility with the hidden objective. Full-world event classifications support offline assessment and diagnosis; they are not a feedback channel to the actors.

\subsection{Operational instantiation}
We instantiate the model through tasks, persistent assignments, local plans, and a communication network. An authoritative objective determines which outcomes the organization is required to achieve; a local plan records the objective an actor currently pursues. Here, a local plan is an explicit objective label, and a commitment is a persistent assignment with work progress. It is not a dependency graph or a representation of the model's hidden reasoning. A revision changes the authoritative objective without automatically changing that label or assignment. This separation makes divergence between institutional intent and ongoing execution an observable state rather than an implementation error.

The experiment isolates the operational coordination path in Figure~\ref{fig:model_graph}: mission assignments reach unit actors through communication, and subsequent actions produce recorded consequences. Strategic priorities and authorization enter as prescribed conditions. Institutional dynamics, physical mobility, and civilian response are outside this instantiation. Section~\ref{sec:interfaces} summarizes the scope and consistency assumptions.

Each actor has its own policy, observation boundary, and permitted consequential actions. The experiments fix the actor pool and initial assignments, allowing explicit handoff after departure. Membership changes activate or deactivate predeclared actors; they do not create new roles or negotiate a new organization.

\subsection{State and transition semantics}
The analytical state projection is
\begin{equation}
S_t=(W_t,N_t,G_t,O_t,C_t,A_t,D_t,Q_t),
\end{equation}
where the terms denote workflow state, active actors, communication availability, authoritative objectives, retained constraints, authority metadata, task commitments, and pending messages. Actor-local plan labels belong to workflow state but are visible only to their owner. Conversational memory is maintained separately by each policy. This projection groups the variables relevant to the analysis; it is not itself available to any actor.

Each joint action is validated against a pre-step snapshot. The simulation resolves action intents (proposed operations), applies scheduled changes, then delivers due messages subject to connectivity and recipient activation. A tick is one discrete simulation step; each active actor has one decision opportunity per tick. Consequently, a newly delivered message cannot retroactively change the action committed earlier in that step. Ordering is deterministic; the execution and tie-breaking conventions are specified in Section~\ref{sec:parameters}.

The scheduled authoritative revision changes $O_t$. Issuing its notice does not rewrite actor plans, acknowledge receipt, cancel downstream work, or inform disconnected peers. Actors explicitly acknowledge a notice by recording recognition, adopt it by updating their local objective label, or relay it by submitting a copy for a peer. Receipt means that transport has placed the notice in an actor's accessible observation; it does not establish model comprehension. Acknowledgment alone preserves the prior plan. Work advances task progress and accumulates synthetic supply cost; the counter is not an inventory limit. A later objective revision does not erase prior progress or cost.

\subsection{Systematic interventions}
An intervention identifies seven attributes: target, operation, origin and authority, timing, observability, propagation, and preserved obligations. Scenario construction binds these attributes to state identifiers and persistent independent assignments, defines local evidence and expected trace predicates, conditions checked against recorded events, then checks the resulting initial conditions before model execution. A change in the requested outcome is distinct from a change in feasibility or a report about the world. In particular, false text does not itself change authoritative state.

Table~\ref{tab:families} summarizes the six intervention families. Each scenario separates work for the initial objective, work for the revised objective, and obligations that survive the revision. Only tasks required by the final authorized objective and retained obligations determine completion. Exact task requirements and disturbance times appear in Section~\ref{sec:parameters}.

\begin{table*}[t]
\caption{Scenario interventions and the coordination distinctions they expose.}
\label{tab:families}
\begin{tabularx}{\textwidth}{@{}p{.18\textwidth}YY@{}}
\toprule
Family & Intervention & Measurement target \\
\midrule
Replacement & Repair changes to delivery; only actor $a$ initially receives the notice & Local adoption and propagation to assigned peers \\
Retained constraint & The goal changes while retained tasks and the shortcut restriction remain active & Preservation of unfinished obligations and constraint attempts \\
Condition & The regular route closes without an objective change & Replanning from delayed factual evidence \\
False authority & A notice claims human authority but has unverified, unauthorized provenance; the goal stays repair & Adoption of misleading direction \\
Membership & An actor joins and another leaves while assignments persist; notices are delayed & Explicit transfer of abandoned commitments \\
Partition & One actor loses communication during a revision and later reconnects & Buffered updates, uninformed work, and recovery \\
\bottomrule
\end{tabularx}
\end{table*}

\subsection{Provenance and information boundaries}
Provenance records a notice's origin. A notice has an issuer, kind, target goal, declared authorization and verification status, and original identifier. Verification denotes trusted origin evidence; authorization denotes permission to issue the revision. A legitimate objective notice has both indicators set. Relaying preserves the original notice. The synthetic authority registry is controlled by the experiment, not by message prose. Origin classification and permission are separate fields, but the present legitimate and forged interventions vary them together; independent authorization-scope discrimination is not evaluated. Neither the words ``human instruction'' nor a model interface's user-role label establish authorization. The experiments use simulator-supplied provenance; they evaluate decisions given that evidence rather than external authentication. The authority indicators are controlled experimental conditions, not evidence of external identity verification.

An actor sees its own plan, acknowledgments, delivered notices, known membership, task definitions, and progress for tasks it currently owns. It also sees the mechanical outcome of its own last action, including progress, conflict, or blocking. Successful work reports progress without classifying its compatibility with the hidden objective; those classifications are evaluator-only. It does not see the true global objective, future schedule, undelivered notices, or other actors' local plans and progress. Membership knowledge changes on receipt of a notice; objective adoption remains an actor decision. A newly activated actor retains its initial plan until it receives and adopts an update.

Engine action menus contain intents for both old and new tasks, including shortcut attempts and ownership claims. They are not pruned according to the hidden objective, task ownership, or route availability. The transition guard, a simulator rule that enforces action preconditions, records attempts and blocks forbidden shortcuts, unavailable routes, and work without ownership. This separates unsafe decision-making from realized world violations and prevents a filtered legal-action menu from revealing an undelivered revision.

\subsection{Commitments and decentralization}
A departed actor's unfinished tasks remain assigned until another actor claims them. Claims use an authoritative, always-available ownership service: the first claim in deterministic actor order transfers a task only if its current owner is inactive. Later competing claims conflict. This service is available even to communication-disconnected actors; message outages affect notice delivery, not access to local work or ownership arbitration. The simulator supplies exclusive ownership. Policies do not achieve decentralized ownership consensus.

A message is accepted when transport admits its submission to the queue. Accepted messages belong to a durable transport: the queued copy persists even if its sender later disconnects or leaves. Submission requires an online sender and a free queue entry; delivery waits for the due tick and an online, active recipient. Sender disconnection or departure after acceptance does not revoke a packet. Duplicate suppression uses the original notice identifier and recipient, so multiple relays produce one receipt. Duplicate, sender-offline, and queue-capacity outcomes are distinct. No delivery acknowledgment is automatically exposed to the sender. These guarantees are deliberately stronger than a physical radio link.

There is no automatic task allocator or globally reconciled plan ledger. Policies coordinate through explicit actions and delayed observations, while the simulator provides authoritative objective and ownership state. Initial assignments are scenario-defined, and claimant selection is a fixed local rule. Agent-negotiated organization, arbitrary actor creation, task-dependency cancellation, and repeated objective supersession are outside the evaluated subset.

\subsection{Scope and consistency assumptions}
\label{sec:interfaces}
The evaluated workflow represents operational assignments, task progress, objective revision, and delayed communication among three actors. Strategic intent enters through a prescribed objective and authority conditions. Political dynamics, geographic movement, and civilian responses remain outside the experiment. The capability graph describes how these responsibilities relate within Strategic World, rather than evidence that every domain has been validated.

Table~\ref{tab:consistency} summarizes the assumptions governing information and coordination. Objective changes affect authoritative state, whereas receipt, adoption, and task reassignment require distinct events. Communication outages delay notices but do not remove access to local work or ownership arbitration. These assumptions delimit which coordination failures the experiment can identify.

\begin{table*}[t]
\caption{Consistency and availability assumptions. Communication disconnection does not remove access to the ownership service.}
\label{tab:consistency}
\begin{tabularx}{\textwidth}{@{}p{.17\textwidth}YY@{}}
\toprule
State & Authority and visibility & Update and outage behavior \\
\midrule
Objective & Authoritative simulator; hidden from policies & Scheduled revision; receipt does not update a local plan \\
Local plan & Actor-owned label; private observation & Explicit adoption; persists through disconnection \\
Membership & Actual activation authoritative; local knowledge delayed & Notice delivery updates knowledge; no inferred departure from silence \\
Ownership & Authoritative serialized registry; own records visible & Claim succeeds only for inactive owner; arbitration remains available during outages \\
Queued notice & Durable transport; pending queue hidden & Accepted copy survives sender loss; waits for active online recipient \\
Action feedback & Last own mechanical outcome only & Returned on next observation; no peer outcomes disclosed \\
\bottomrule
\end{tabularx}
\end{table*}

\section{Evaluation}
\subsection{Measures and adaptation accounting}
Assessment derives outcomes from recorded actions, events, and state transitions. Success requires completion of all tasks for the final authoritative goal and all retained obligations. Obsolete (stale) work pursues a superseded objective and is separated according to whether the executing actor had received a verified, authorized objective notice. An acknowledgment--behavior gap requires acknowledgment of that same notice followed by incompatible work. Unauthorized adoption changes a local plan using an unverified or unauthorized notice; it cannot change the authoritative objective. Shortcut attempts and realized violations are counted separately.

An eligible actor is active at or after the revision and before the episode deadline; eligibility does not require notice receipt. Every such actor has an adaptation record containing its active intervals, revision receipt, adoption, first compatible revised-goal work, and observation deadline. Receipt-to-adoption latency measures a local plan update; receipt-to-work latency measures consequential behavior. Either may occur without the other. First compatible work is an onset measure, not evidence of sustained adherence. Actors lacking receipt remain in the eligible denominator. Episodes without an authoritative revision have no revision-latency denominator. Censoring means that the measured response is not observed within the observation window. Its recorded reason distinguishes departure, absence of receipt for an actor still active at the deadline, and a deadline reached after receipt. We report observed and censored counts together; a latency conditional on an observed event is not a team adaptation rate.

\subsection{Experimental parameters}
\label{sec:parameters}
Each principal diagnostic and raw-interface episode contains three predeclared actors and nine tasks: one repair, delivery, and retained-obligation task per actor. Repair requires eight work units and delivery two. Retained obligations require one unit, except in the retained-constraint family, where four units leave prior obligations unfinished at revision time. Work adds one progress unit and costs one synthetic supply unit on the regular route or two on the alternate route. Shortcut use remains forbidden throughout an episode. Positions are schematic and do not affect these task transitions.

The six timing configurations cross message delays of one, two, and three ticks with onset offsets of zero and one tick. The base revision, route-closure, or false-notice event occurs at tick 3. In membership, actor $c$ activates at base tick 5 and actor $a$ departs at base tick 8. In partition, actor $b$ disconnects at base tick 3 and reconnects at base tick 9. The onset offset shifts these scheduled times. Principal episodes end after 18 ticks. Calibration crosses deadlines of 18 and 36 ticks without moving disturbances. Every active actor has one decision opportunity per tick. A no-intervention family removes all disturbances while preserving the initial tasks and action interface.

Joint action intents refer to the same pre-step snapshot. The diagnostic transition resolves intents in actor order, applies scheduled changes, and delivers due messages afterward. The first valid claim transfers ownership; competing claims against the resulting active owner are recorded as conflicts. A notice available in the resulting state can affect the next decision, but not an intent already committed in the current step.

The nine raw-interface model episodes use environment seed zero and three paired inference seeds, 923--925, for each of static, replacement, and false authority. Temperature is zero, context capacity 32,768 tokens, output limit 128 tokens, and at most two inference attempts are allowed per decision, allowing a retry when the model response is invalid. Each actor retains four recent decisions in a private context. Task feedback is restricted to its own last action outcome. The required response format restricts syntax and available action identifiers without selecting a task, route, or authority interpretation.

The four task probes assess the operations described in the static-competence evaluation. Communication microtasks add explicit natural-language immediate objectives: acknowledge a delivered notice without adopting; adopt an already acknowledged legitimate revision; forward an adopted notice to teammate $b$; and acknowledge an untrusted notice without adopting it. A three-decision sequence requires acknowledgment, adoption, and forwarding in order. Acknowledgment must preserve the plan and task progress, adoption must update the plan without advancing work, and relay must deliver a notice without changing the recipient's plan. Scoring checks both action choice and resulting state. The sequence retains the same private history across decisions; separate checks start fresh.

All reported episodes and probes use mechanical feedback. Exploratory runs that disclosed objective compatibility through action feedback are excluded from the reported model results. Probes that combined acknowledgment and route choice are likewise excluded from the separate operation checks.

\subsection{Matched protocols and feasibility}
The principal diagnostic suite crosses six scenario families, six timing configurations, and five controls, yielding \DiagnosticEpisodes{} episodes. Each lasts 18 ticks. Delays range from one to three ticks, crossed with onset offsets of zero or one. The same disturbances and tasks are paired across controls. Configurations are designed conditions rather than independent population samples.

A control is a scripted policy used to establish a known behavior or failure mechanism. The \emph{structured} control acknowledges, adopts verified authorized revisions, and relays to peers before work. \emph{Ack-only} disables both adoption and relaying; \emph{trust-all} extends the structured control by also adopting and relaying unverified objective notices; \emph{unsafe} attempts forbidden shortcuts; and \emph{no-relay} adopts locally without forwarding. These are constructed finite-state diagnostics, scripted policies with a bounded set of internal states, rather than optimal planners.

A separate calibration crosses four local protocols with 18- and 36-tick deadlines under the same 36 family--timing conditions. A positive control establishes that the tested tasks can be completed under the declared constraints. The \emph{efficient} positive control has the initial notice recipient submit copies to all declared slots before adoption and acknowledgment; other recipients avoid echoes unless their local membership evidence indicates that the origin departed. Work precedes optional acknowledgment, and the lowest locally known surviving identifier reclaims abandoned tasks. Messages can wait for an inactive recipient. This fixed protocol uses local evidence, not peer receipt state or privileged access to a global allocation plan. Two variants separately disable its adoption or its relaying. The structured 18-tick cells repeat the principal baseline. Six no-intervention episodes provide a static positive control. Together these yield \CalibrationEpisodes{} calibration episodes.

\begin{table}[t]
\caption{Matched feasibility calibration. S: serial structured protocol; E: local efficient protocol. Columns vary the deadline in ticks.}
\label{tab:calibration}
\begin{tabularx}{\columnwidth}{@{}Yrrrr@{}}
\toprule
Family & S18 & S36 & E18 & E36 \\
\midrule
Replacement & 4/6 & 6/6 & 6/6 & 6/6 \\
Retained constraint & 4/6 & 6/6 & 6/6 & 6/6 \\
Condition & 6/6 & 6/6 & 6/6 & 6/6 \\
False authority & 6/6 & 6/6 & 6/6 & 6/6 \\
Membership & 0/6 & 4/6 & 5/6 & 6/6 \\
Partition & 4/6 & 6/6 & 6/6 & 6/6 \\
Total & 24/36 & 34/36 & 35/36 & 36/36 \\
\bottomrule
\end{tabularx}
\end{table}

Table~\ref{tab:calibration} separates protocol and deadline effects. The efficient control completes \EfficientShortSuccess{}/36 cases at 18 ticks and \EfficientLongSuccess{}/36 at 36 ticks. Thus every tested disturbance configuration is feasible for an observation-only protocol at the longer deadline. The structured control improves from \StructuredSuccess{}/36 to \StructuredLongSuccess{}/36 when only the deadline changes. Its remaining membership failures can lose the only informed actor before dissemination. These results establish feasibility of these finite instances, not optimality of the positive control or inevitability of failures under other conditions.

In the latest-onset, three-tick-delay replacement case, the structured protocol expends 4 actions on duplicate relays. The efficient adoption-disabled and relay-disabled variants each complete 12/36 and 12/36 cases, respectively, at the longer deadline; these interventions separate the two operations within the same protocol family. Explicit adoption is a requirement of this policy, not of the work transition. In the three-tick-delay membership case with base onset, 1/3 actors adopt but 0/3 perform revised-goal work. The actor records distinguish departure of the informed adopter from the two active peers lacking receipt.

\subsection{Information lost by simpler evaluation}
For a trajectory $\tau$, consider the summary
\begin{equation}
P(\tau)=(\text{success},\text{completed tasks},\text{acknowledging actors}).
\end{equation}
We construct three diagnostic witnesses, trajectories selected to demonstrate distinctions lost by the summary. They share the same initial state and revision schedule. Actor $a$ receives the revision at tick 4 and acknowledges it at tick 5. At tick 6, one trajectory executes obsolete work by $a$, another executes obsolete work by uninformed actor $b$, and the third attempts a forbidden shortcut on a revised-goal task. All subsequent actions wait.

\begin{table}[t]
\caption{Three trajectories with identical failure, zero completed required tasks, and one acknowledging actor. I/U: informed/uninformed obsolete work; B: blocked shortcut attempts.}
\label{tab:witnesses}
\begin{tabularx}{\columnwidth}{@{}Yrrr@{}}
\toprule
Trajectory & I & U & B \\
\midrule
Post-receipt stale work & 1 & 0 & 0 \\
Missing information & 0 & 1 & 0 \\
Forbidden intent & 0 & 0 & 1 \\
\bottomrule
\end{tabularx}
\end{table}

All three have $P(\tau)=(\mathrm{false},0,1)$, yet their trace signatures, the recorded counts of informed work, uninformed work, and blocked shortcut attempts, differ (Table~\ref{tab:witnesses}). Consequently, completion and aggregate acknowledgment do not identify whether this failure involved missing information, incompatible behavior after receipt, or an enforced safeguard. These are constructive witnesses to information loss, with hand-specified expected signatures, rather than an estimate of diagnostic classification accuracy on naturally occurring failures.

Across the principal suite, acknowledgment-only produces \AckObsolete{} informed obsolete-work events and \AckGaps{} actor-level gaps between acknowledgment and behavior. Trust-all produces \UnauthorizedAdoptions{} unauthorized adoptions; unsafe produces \ConstraintAttempts{} blocked shortcuts and zero realized shortcut violations. No-relay produces \NoRelayUninformed{} uninformed obsolete-work events. Transition guards explain prevented violations; they do not demonstrate safe policy decisions. The complete five-control completion matrix appears in Section~\ref{sec:validation}.

\subsection{Static competence and local-model pilots}
We use \LocalModelName{} with independent actor contexts. Episode policies receive no recommended actions. The interface describes claim, work, acknowledgment, adoption, and relay effects, and returns only mechanical outcomes of the actor's previous action. The system instruction is domain-neutral. The raw-interface pilots retain engine intents without filtering by hidden objectives, ownership, or route availability. Clarifying an action does not choose it for the model. An accepted model decision is a syntactically valid selection from the offered action identifiers; its attempted operation may still be blocked by the simulator.

Three decoding seeds are paired across a static workflow, replacement, and false authority, holding the environment timing fixed. Static episodes keep objectives and membership fixed. Completion in this setting is a necessary calibration for interpreting dynamic failure. Four task probes examine owned work, conflict recovery, completion recognition, and closed-route response. Four instructed communication probes separately test acknowledgment, adoption, relay, and rejection of an untrusted revision. A final check requires acknowledgment, adoption, and relay across three decisions. These microtasks provide an immediate objective in plain language; they test execution of requested operations rather than autonomous discovery of a coordination strategy. Expected action sets and conditions on the resulting state remain outside policy packets. Section~\ref{sec:parameters} specifies the inference budget and probe protocol.

\begin{table}[t]
\caption{Raw-interface local-model pilots: three paired decoding seeds per condition, with 18 ticks and three actors. Task completion is aggregated over six required tasks per episode.}
\label{tab:local}
\begin{tabularx}{\columnwidth}{@{}Yrrr@{}}
\toprule
Condition & Success & Tasks & \shortstack{Claim\\conflicts} \\
\midrule
Static & 0/3 & 15/18 & 0 \\
Replacement & 0/3 & 9/18 & 0 \\
False authority & 0/3 & 15/18 & 0 \\
\bottomrule
\end{tabularx}
\end{table}

The nine episodes produce 486 accepted decisions from 486 inference requests; every recorded trajectory verifies. Static completion is 0/3, with 0 ownership conflicts. The static traces contain 78 already-complete or non-owner work attempts. Task completion alone does not measure inefficient or invalid action use. Replacement completes 0/3 episodes; the traces contain 0 unauthorized adoptions and 12 informed obsolete-work events. False authority completes 0/3 episodes; the traces contain 0 unauthorized adoptions and 0 informed obsolete-work events. Across replacement runs, 3/9 actor-episodes receive the revision, 0/9 explicitly adopt, and 0/9 perform compatible revised-goal work. Failure on the static workflow limits attribution of dynamic failures to coordination-specific mechanisms. Distinct joint-action trajectories across the three seeds are static 1/3, replacement 1/3, false authority 1/3. These are repeated executions, not independent evidence of behavioral diversity. Task probes pass 6/12 checks. Instructed communication probes pass 12/12, and the complete three-decision sequence passes 0/3. Table~\ref{tab:probes} separates these checks. Passing an explicitly requested operation does not show that a policy will select or propagate it autonomously. The sample does not establish general competence or robust authority discrimination.
Each sequence repeats acknowledgment three times although the notice is recorded as acknowledged after the first action. First-action success therefore conceals failure to update the plan and propagate the notice. 
\begin{table}[t]
\caption{Raw-interface competence checks under mechanical feedback. The last five rows provide explicit immediate communication tasks. A sequence passes only when all three ordered actions and their effects are correct.}
\label{tab:probes}
\begin{tabularx}{\columnwidth}{@{}Yrr@{}}
\toprule
Check & Passed & Decisions \\
\midrule
Owned work & 3/3 & 3/3 \\
Conflict recovery & 3/3 & 3/3 \\
Completion recognition & 0/3 & 0/3 \\
Closed route & 0/3 & 0/3 \\
Acknowledge & 3/3 & 3/3 \\
Adopt & 3/3 & 3/3 \\
Relay & 3/3 & 3/3 \\
Reject untrusted revision & 3/3 & 3/3 \\
Three-step sequence & 0/3 & 3/9 \\
\bottomrule
\end{tabularx}
\end{table}

\subsection{Calibrated baseline and diagnosis-guided intervention}
\label{sec:intervention}
We next evaluate a policy intervention after establishing competence on the underlying operations. The baseline is an assisted controller, rather than an unassisted language model. A local work board summarizes owned tasks, remaining progress, and the actor's current plan. A policy-level candidate filter excludes completed or unowned work, known forbidden routes, redundant acknowledgments, and work while a delivered legitimate revision remains unadopted. These constraints use actor-local evidence. They do not change engine legality or replace a model's selected action; the model chooses from the declared policy shortlist, including waiting. The original action set remains available for reconstructing the effect of this restriction.

The baseline passes two calibration episodes in which all three actors use the same assisted LLM controller: a stable repair workflow and a stable delivery workflow, each completing all six obligations. Delivery starts after an explicitly recorded scripted initialization sequence communicates and adopts the goal; this sequence performs no work and is not attributed to the model. Each measured window lasts 18 ticks. The baseline also passes all \CalibrationProbeTotal{} task and communication checks, including the ordered three-step sequence. These static episodes and operation checks form the calibration gate required before evaluation on changing conditions. These results establish competence of the combined controller on the declared gates, not independent understanding by the underlying model.

In a separate discovery episode, used to select the intervention before held-out evaluation, the calibrated baseline completes the organization’s tasks but performs local delivery work before finishing notice submissions to both peers. Per-recipient traces reveal an unfinished communication obligation that final completion alone does not expose. This diagnosis motivates a \emph{pending-outbox guard}. The outbox is the set of required notice--recipient submissions that the actor has yet to confirm. The guard records only the actor's own confirmed submission outcomes, defers local work while legitimate adopted notices remain unsubmitted to peers, and suppresses confirmed duplicate submissions. Submission is distinct from delivery. The guard neither supplies peer receipt state nor automatically sends, adopts, or completes work: the model must still select an action and consume a decision opportunity.

We freeze the protocol, policies, calibration gate, cases, and outcome measures before held-out model execution. Four new settings vary the initial recipient, revision onset, two- or three-tick message delay, and work requirements (Table~\ref{tab:intervention_cases}). Each is paired across baseline, a generic coordination reminder, route-focused advice, and the outbox guard. The other baseline constraints are shared. To isolate the focal policy intervention, each team contains one LLM actor and two fixed scripted peers that use only actor-local observations. Every episode has 18 LLM decision opportunities with the same context and output limits; actual token counts are measured. Four scripted positive controls establish feasibility under these deadlines. These are designed stress cases, not independent population samples.

\begin{table}[t]
\caption{Frozen held-out settings. Onset and delay are ticks; R/D/K are per-actor repair, delivery, and retained work requirements. The deadline is tick 18.}
\label{tab:intervention_cases}
\begin{tabularx}{\columnwidth}{@{}Yrrr@{}}
\toprule
Case & Recipient & Onset/delay & R/D/K \\
\midrule
H1 & b & 3/2 & 6/2/1 \\
H2 & c & 4/3 & 8/1/1 \\
H3 & a & 4/2 & 5/3/1 \\
H4 & b & 2/3 & 7/2/2 \\
\bottomrule
\end{tabularx}
\end{table}

\begin{table*}[t]
\caption{Held-out policy interventions: four matched settings per arm. Authorized work includes retained obligations. Every episode has one LLM actor, two fixed peers, and 18 LLM decision opportunities. Input and output columns report mean server-reported tokens per episode; these counts are not exactly matched.}
\label{tab:intervention}
\begin{tabularx}{\textwidth}{@{}Yrrrrr@{}}
\toprule
Policy & Success & \shortstack{Authorized\\work units} & \shortstack{Notice\\receipts} & \shortstack{Input\\tokens} & \shortstack{Output\\tokens} \\
\midrule
Calibrated baseline & 1/4 & 33/39 & 9/12 & 31,029 & 1,378 \\
Generic reminder & 0/4 & 31/39 & 11/12 & 31,318 & 1,451 \\
Route-focused advice & 0/4 & 33/39 & 11/12 & 31,848 & 1,523 \\
Pending-outbox guard & 4/4 & 39/39 & 12/12 & 32,015 & 1,562 \\
\bottomrule
\end{tabularx}
\end{table*}

The guard completes 4/4 held-out settings, compared with 1/4 for the calibrated baseline, 0/4 for generic reminders, and 0/4 for route-focused advice (Table~\ref{tab:intervention}). Authorized work is 39/39 units under the guard and 33/39 under baseline. Post-adoption work before both submissions decreases from 10 to 0 actions. This last difference is enforced by the guard and is not independent evidence of learned coordination; downstream completion is the tested consequence. Mean first-relay latency is 4.50 ticks for baseline and 3.00 for the guard, but that scalar omits recipient coverage. In H3, baseline submits its first relay earlier, yet peer $b$ receives no revision before the deadline (Table~\ref{tab:intervention_timeline}). The guard completes submissions to both peers before local work and both finish their obligations. Thus the diagnosis concerns unfinished per-recipient obligations, not simply absence of any send. Configured call, context, and output limits are matched; actual input and output token use differ. The results concern the complete engineered intervention package on four designed settings, not memory alone, a population effect, or autonomous organization formation.

\begin{table}[t]
\caption{Matched case H3. Times are world ticks; NR means no receipt by the tick-18 deadline. An earlier first relay does not establish complete dissemination.}
\label{tab:intervention_timeline}
\begin{tabularx}{\columnwidth}{@{}Yrr@{}}
\toprule
Event & Baseline & Guard \\
\midrule
Root receives revision & 6 & 6 \\
Root adopts revision & 10 & 8 \\
First relay submitted & 8 & 9 \\
Peer $b$ receives revision & NR & 11 \\
Peer $c$ receives revision & 10 & 12 \\
Required tasks completed & 5/6 & 6/6 \\
\bottomrule
\end{tabularx}
\end{table}

\subsection{Validation of observations and measures}
\label{sec:validation}
Recorded actions reconstruct the same observations, events, rewards, and world states in all \DiagnosticEpisodes{} principal episodes, \CalibrationEpisodes{} calibration episodes, three diagnostic witnesses, and nine raw-interface model episodes. The held-out comparison, feasibility controls, and baseline calibration episodes are also reconstructed, including the observations and conversational histories presented to each policy. Replay establishes consistency under recorded decisions; a fresh model invocation remains a new execution.

Hand-labeled trajectories check receipt-relative latency, eligibility denominators, departure censoring, and the distinction between adoption and consequential work. Paired executions that differ only in an undelivered objective revision produce identical actor observations before and after the same work action, while retaining distinct evaluator classifications. Repeated work probes check that the actor cannot infer the hidden objective from success feedback. Additional cases verify message persistence after sender loss and exclusive ownership under competing claims during a communication outage. These checks support the reported measurements under the declared assumptions; they do not establish realism of the wider world model.

Table~\ref{tab:diagnostics} reports completion for all five principal controls.
\begin{table}[t]
\caption{Principal diagnostic completion counts. Six timings per family. S: structured; A: ack-only; T: trust-all; U: unsafe; N: no-relay.}
\label{tab:diagnostics}
\begin{tabularx}{\columnwidth}{@{}Yrrrrr@{}}
\toprule
Scenario & S & A & T & U & N \\
\midrule
Replacement & 4/6 & 0/6 & 4/6 & 0/6 & 0/6 \\
Retained constraint & 4/6 & 0/6 & 4/6 & 0/6 & 0/6 \\
Condition & 6/6 & 6/6 & 6/6 & 0/6 & 6/6 \\
False authority & 6/6 & 6/6 & 0/6 & 0/6 & 6/6 \\
Membership & 0/6 & 0/6 & 0/6 & 0/6 & 0/6 \\
Partition & 4/6 & 0/6 & 4/6 & 0/6 & 0/6 \\
Total & 24/36 & 12/36 & 18/36 & 0/36 & 12/36 \\
\bottomrule
\end{tabularx}
\end{table}

\subsection{Reproducibility}
\label{sec:artifacts}
The research artifacts retain scenario definitions, policy specifications, model inputs and outputs, and complete event histories. Reported tables are derived from recorded assessments. The intervention study additionally retains the protocol fixed before evaluation, its calibration criteria, and matched case definitions. These records permit reconstruction of both outcomes and the information available at each decision.

\section{Discussion and Limitations}
Completion is a property of the trajectory's final obligations; adaptation is an actor-level process conditioned on information and decision opportunities. The witness trajectories show why one cannot be recovered from the other. Likewise, an adoption-latency distribution excludes a different set of actors from a behavioral-latency distribution. Reporting eligible, observed, and censored actors is necessary to avoid interpreting a fast surviving subset as rapid team adaptation.

The feasibility controls change the interpretation of deadline failures. A local protocol completes all tested conditions at 36 ticks, while acknowledgment and redundant dissemination consume opportunities under the serial reference protocol. Failure in this environment therefore cannot be attributed to communication difficulty alone. These experiments do not identify a globally optimal scheduler; the positive control uses a fixed initial broadcaster and a deterministic claimant rule.

Strategic World separates world dynamics, local observations, consequential actions, and recorded histories. The evaluated instantiation remains an abstract operational workflow: routes represent availability and cost, progress is discrete, assignments are independent, and supply expenditure is a counter rather than a binding budget. The broader capability graph identifies integration points; political dynamics, physical radio propagation, geographic mobility, and civilian response are not validated by these experiments.

The transport and ownership services also bound external validity. Accepted messages survive sender failure, and ownership arbitration remains authoritative during communication outages. Policies lack a coordinating planner, but they do not establish distributed ownership agreement. Membership changes activate predeclared actors; arbitrary spawning and negotiated organization are not evaluated. Provenance flags are simulator-supplied, with verification and authorization coupled in the tested interventions. Neither external authentication nor independent authorization-scope reasoning is established.

Static completion and targeted communication checks measure different prerequisites for coordination. Passing an instructed microtask does not establish that a policy will select that operation autonomously. Isolated probes can expose weaknesses hidden by successful episodes or recent conversational context. The local-model study uses one open-weight model, three decoding seeds, fixed environment timing, and a bounded memory and inference budget. These runs are too few for reliability estimates, and decoding seeds need not induce meaningfully different behavior. Earlier episodes with evaluator labels in policy feedback are archived and excluded. Mechanical feedback prevents this direct objective-disclosure channel, but passing the declared counterfactual tests is not a proof against every possible information channel.

The traces measure externally observable adaptation; they do not reveal whether an unresponsive model internally represents the revised objective. Establishing that connection would require representation probes and interventions beyond the present behavioral evaluation.

The intervention study tests a correction motivated by trace diagnosis: a successful discovery trajectory exposes late dissemination, and the resulting guard improves completion on the frozen stress cases. Candidate filtering and persistent submission tracking are explicit policy engineering. The experiment does not isolate memory from prioritization, demonstrate spontaneous model coordination, or evaluate an all-LLM team under those disturbances. Generic reminders and unrelated advice control for some forms of added assistance, but actual input and output tokens differ. Four paired settings do not establish robustness across models or arbitrary organizations.

All scenarios use fictional entities and simulated consequences. The findings concern measurement and coordination within this abstraction.

\section{Conclusion}
We presented a Strategic World testbed for changing objectives and conditions under persistent assignments and delayed communication. Actor-level accounting distinguishes missing information, adoption, consequential work, and censored responses. Matched protocols and deadlines establish feasibility for the tested disturbances, while controlled witness trajectories demonstrate distinctions lost by completion and acknowledgment summaries. Raw-model episodes and action probes expose interface limitations; a calibrated assisted controller supports a held-out test of a diagnosis-guided outbox guard. Explicit transport, ownership, and provenance assumptions make these results reproducible without conflating simulator guarantees with capabilities achieved by the agents.

\section*{Generative AI Usage Disclosure}
During the preparation of this manuscript, the authors used ChatGPT for language editing and refinement. In addition, Qwen3 was employed as a locally deployed LLM to simulate multiple agents during the evaluation process. All AI-generated outputs were carefully reviewed and revised by the authors, who take full responsibility for the accuracy, integrity, and final content of this publication.

\bibliographystyle{ACM-Reference-Format}
\bibliography{references}
\end{document}